# Observation of galactic cosmic ray spallation events from the *SoHO* mission 20-Year operation of LASCO

S. Koutchmy [1] * , E. Tavabi[1,2] & O. Urtado[1]

[1]*Institut dAstrophysique de Paris, UMR 7095, Sorbonne Universite- CNRS and UPMC, 98 Bis Bd. Arago, 75014 Paris, France.*
[2] *Physics Department, Payame Noor University (PNU), 19395-3697, Tehran, I. R. of Iran.*

**ABSTRACT**

A shower of secondary Cosmic Ray (CR) particles is produced at high altitudes in the Earth's atmosphere, so the primordial Galactic Cosmic Rays (GCRs) are never directly measured outside the Earth magnetosphere and atmosphere. They approach the Earth and other planets in the complex pattern of rigidity's dependence, generally excluded by the magnetosphere. GCRs revealed by images of single nuclear reactions also called spallation events are described here. Such an event was seen on Nov. 29, 2015 using a unique LASCO C3 space coronagraph routine image taken during the *Solar and Heliospheric Observatory (SoHO)* mission observing uninterruptedly at the Lagrangian L1 point. The spallation signature of a GCR identified well outside the Earth's magnetosphere is obtained for the 1$^{st}$ time. The resulting image includes different diverging linear "tracks" of varying intensity, leading to a single pixel; this frame identifies the site on the silicon CCD chip of the coronagraph camera. There was no solar flare reported at that time, nor Coronal Mass Ejection (CME) and no evidence of optical debris around the spacecraft. More examples of smaller CR events have been discovered through the 20 years of continuous observations from *SoHO*. This is the first spallation event from a CR, recorded outside the Earth's magnetosphere. We evaluate the probable energy of these events suggesting a plausible galactic source.

## 1 INTRODUCTION

The CCD imaging instruments of the *SoHO* mission (Domingo et al. 1995) of ESA and NASA, including the Large Angle and Spectrometric Coronagraphs (LASCO) and the Extreme ultraviolet Imaging Telescope (EIT) imagers, are sensitive to Solar Energetic Particles (SEP) in the MeV up to GeV range. Yagoda (1962), Obayashi (1964) and Roederer (1964) reported that a myriad of impacts is continually recorded at the time of big flares and CMEs. Higher energy particles in the GeV and in the more energetic range up to $10^{21}$ *eV* are today called cosmic rays (CR), *e.g.,* see Freier et al. (1948); Chandrasekhar and Fermi (1953); Fermi (1954); Gaisser (1990) and Dorman (2006), they are primarily made of protons. Usually remotely registered at ground-based (G-B) observatories by different methods, following their interaction in the upper atmosphere, at a column depth of order 1033 g/cm$^2$ (Kudela 2009). They produce a shower of secondary particles as a result of a sequence of reactions of the primary CR particle. In G-B observations they are also registered in situ with the neutron monitors preferably at high altitude sites to be closer to the primary site of CRs interaction with the Earth atmosphere but G-B observations exist at sea level including the large facilities developed for analyzing the very high energies (Smart and Shea 2009).

The primary particles have been deflected by the galactic magnetic field making their trajectory in the interstellar space impossible to follow (Baade and Zwicky 1934; Butt 2009). These particles are then deviated in the much stronger magnetic field of the heliosphere and finally by the magnetic field of our magnetosphere (Dorman 2006; Kahler 1992; Smart and Shea 2009; Miroshnichenko 2015), as illustrated in Fig. 1. Known and studied for one century and despite the existence of a very extended scientific literature, the definite origin of these particles is still not established because of the difficulty to identify the source(s) in the sky and to theorize the mechanisms for producing such extreme energies. The popular and classical suggested scenario is the so-called Fermi mechanism (Chandrasekhar and Fermi 1953; Fermi 1954) of acceleration of elementary particles in the magnetized shock front of the super-novae explosions occurring in our galaxy, including the remnants. Neutron star activity, super-flares on O-type massive stars and other exotic stellar phenomena are today also considered. Even more extremely energetic CRs are seldom registered and new sources related to black hole "activity" are suggested from our own galactic center and more probably, from extra-galactic objects (Butt 2009).

* E-mail: koutchmy@iap.fr

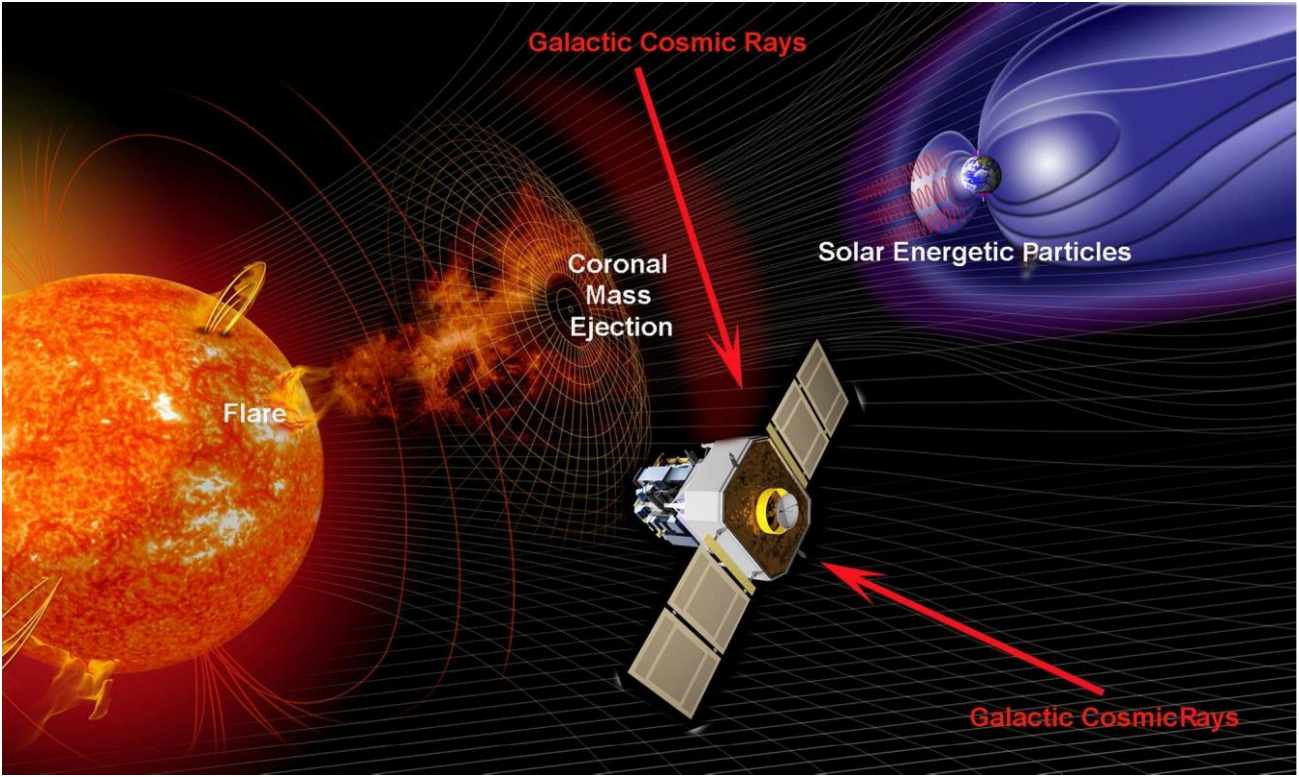

**Figure 1.** Synoptic view of the heliosphere and magnetosphere with magnetic structures, schematically showing the position of the *SoHO* spacecraft with energetic particles of different origins coming in (frames from movies by NASA and ESA are used to consider this synoptic image).

A very popular and well-established quantitative parameter of CR is shown by the display of their energy spectrum which shows the flux of CR versus the energy of the particles as a quasi-power law (Kudela 2009) for the parts corresponding to energies $E > 1$ GeV. From this brief presentation it is clear that any new diagnostic of galactic CRs will be welcome especially from observations made in space well outside our magnetosphere. This has been possible thanks to uninterrupted observations performed by the *SoHO* spacecraft launched on 1995 Dec. 02 and still in operation today as far as the coronal imaging experiment is concerned. The spacecraft is put in a halo orbit of the Lagrangian L1 point situated roughly 0.01 AU interior to the orbit of the Earth system, see Fig. 1.

Since they are not shielded by the Earth's magnetosphere, *SoHO* suffered damage from CRs (Curdt and Fleck 2015), starting with the solar array degradation and the solid state recorder system. From an analysis of the upsets in the recorder system for solar cycle 23 (1996 to 2008) it was found (Curdt and Fleck 2015) that 94% of effects were due to CR of galactic origin (GCRs). However the Forbush effect as known for a long time from the neutron monitor recording (Simpson 1957; Lara et al. 2005) was not observed by Curdt and Fleck. Still *SoHO* imaging experiments using a CCD chip as a detector produced a myriad of dots and tracks on LASCO coronal images (Domingo et al. 1995) and many movies showing the "snowstorm" effect produced by Solar Energetic Particles (SEPs) resulting from major flares and from large-scale Coronal Mass Ejections CMEs (*e.g.* Cane et al. 2010; Kahler 1992) see Fig. 2. We concentrate here on images coming from the C3 externally occulted coronagraph (Brueckner et al. 1995) that produces a 16° large field-of-view (FOV) around the Sun. Images show i) the solar corona plasma structure; ii) the large halo of the dusty F-corona surrounding the Sun (Koutchmy and Lamy 1985); iii) bright stars, sun-grazing comets and planets crossing the FOV;

iv) images of space debris; v) bright dots and linear tracks mainly produced by SEPs and presumably, by GCRs more easy to record when the Sun is quiet.

## 2 OBSERVATIONS

Fig. 1 presents a synoptic view of the solar system showing the spacecraft *SoHO* with the different components identified. Note that space debris produce rather out of focus optical effects over the optical FOV. They are affected by the internal vigneting and masking, including the effect of the external occulters; images of their tracks correspond to extended objects moving rapidly across the FOV easy to identify. This is in contrast with the effects of SEP and CRs that produce pixel-width signature (bright dots and tracks) over the 20 x 20 mm$^2$ chip of the CCD camera, everywhere over its surface including the part optically masked by the external occulter. This well-known effect is prominent at the time of big X-type flares on the Sun and it is easily evaluated from the processed images and movies produced by the *SoHO's* LASCO experiment and stored in different data bases at NASA, ESA and the participating labs, see Fig. 2. We refer to observations of the classical and best studied X 5.7 flare event of July, 14, 2000 (Bastille day flare) to extract some relevant specifications of the event with respect to CRs (Klein et al. 2001; Belov et al. 2001; Mishev and Usoskin 2016). Note that the flare produced (Lara et al. 2005) a definite Ground Level Enhancement (GLE) # 59 in Neutron Monitors (NMs). CRs of energies up to 6 GeV are reported from Neutron Monitors recording (Belov et al. 2001). The occurrence of millions of point-like impacts with a few tracks after a solar flare is an important feature of the CCD images that permits a definite identification of the tracks produced by the ionization effect of a SEP inside the Si plate of thickness of order of 60 microns.
Images also show near the surface of the Sun million degrees hot solar plasma cooling down while suspended in an arcade of postflare magnetic loops. SEPs have different origins. Primary SEP are produced in the reconnection region of the flare itself at the solar surface with impulsive events. Even more significant secondary SEP are produced as the result of interaction with the ejected shock front of the CMEs (gradual events, Klein et al. 2001; Curdt and Fleck 2015; Firoz et al. 2011). For the sake of simplicity planets and bright streaks due to debris are not shown as Fig. 1 although they are well recorded by the LASCO coronagraphs. The trajectory of a GCR penetrating inside the solar system is shown at left with a red color: it is shown exaggeratedly being deviated by the heliospheric magnetic-field before reaching the spacecraft (not to scale). Tracks are produced when the trajectory of the particle is close to parallel to the surface of the CCD. We note that the cross section of the nucleus of the silicon atoms is similar to the cross section for Al and other metallic components of the spacecraft; it makes a convenient detector of energetic CRs chip in case of a collision inside the chip. Such interactions also appear at times of low or no solar activity (no trace of flaring active region on the disk, no GOES soft X-ray signatures) and no CME event.

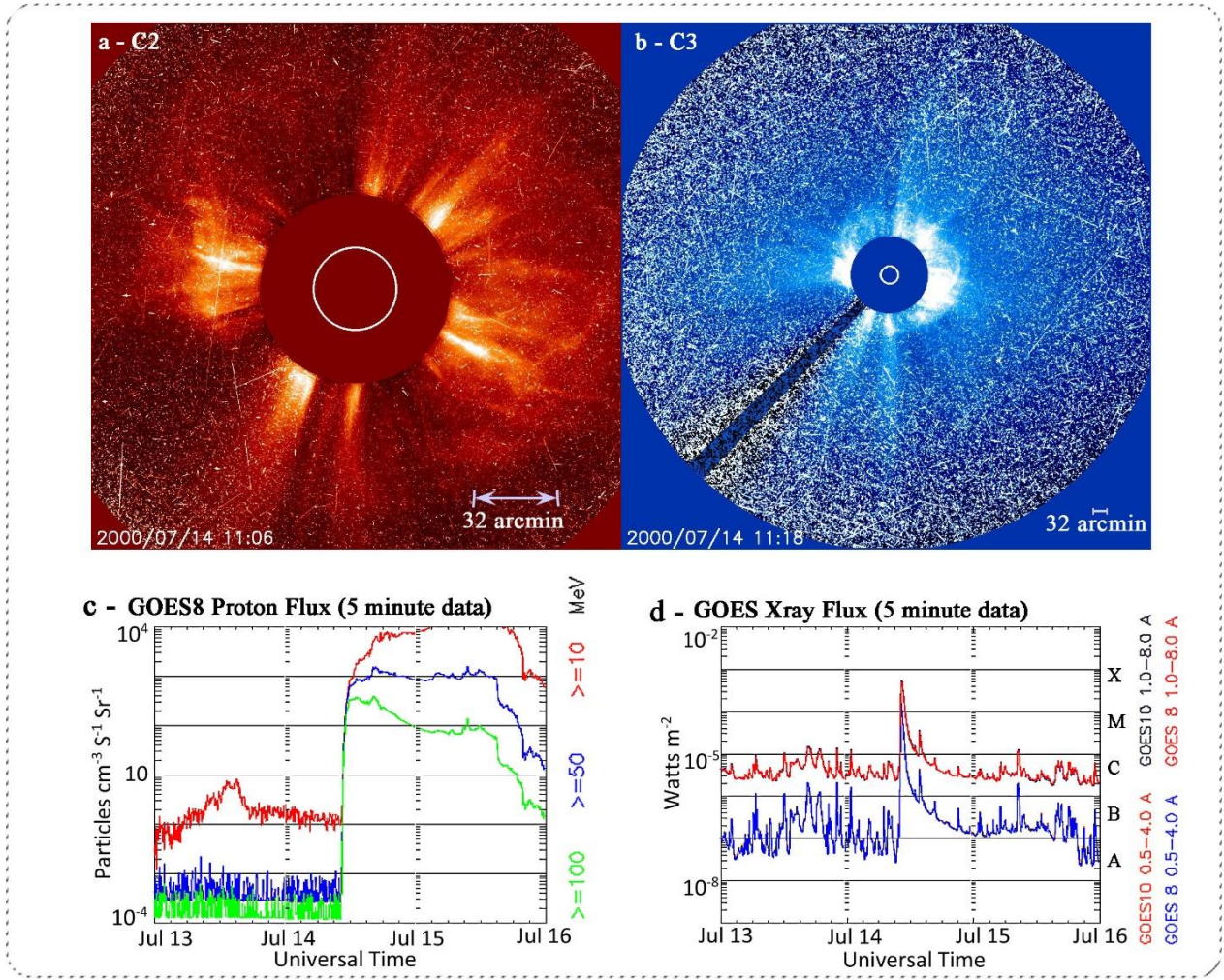

**Figure 2.** Bastille day extreme solar flare (10:30 U.T., 14[th] July 2000) of active region 9077 to show (a & b) the effect of energetic particles recorded inside the CCD chip using the LASCO imaging instrumentation of the *SoHO* spacecraft observing the Sun from the L1 Lagrangian point. The event corresponds to a massive X-type flare as demonstrated by the sudden increase of GOES8 proton flux (c) and the GOES X-ray flux (d).

The GCRs usually produce a signature not different from the dot or line signatures given by SEPs. Their energy (Kudela 2009) is distributed according to a power law function which is well established from decade long observation at ground, after spallation nuclear reactions with the high Earth atmosphere atoms. In the example of the 2000 Bastille day extreme large event, we tried, without success, to look at the more complex signature as the "star" image described by Levi Setti and Tomasini (1952), in LASCO images, especially frames taken after the X- ray flare when the most energetic protons hit the CCD chip. We repeated the search, again without success, for the recent "double" X-type flare of Sept. 6 (disk event) and of Sept. 9 (limb event), 2017 using both the sequences from the C2 and the C3 LASCO coronagraphs of *SoHO*. At presumably higher energy (see Fig. 3), a spallation event see *e.g.* Kowalczyk (2008) and Krasa (2010) and pictures of what was called "star" from the former appearance in photographic plates (Levi Setti and Tomasini 1952) may indeed result in a photographic emulsion, see Yagoda (1962). This is exactly what was noticed during an examination of a movie made from routinely processed C3/LASCO coronagraph observations of the *SoHO* mission, at 11:30 UTC of 29 Nov. 2015, a time of very low solar activity on the Sun.

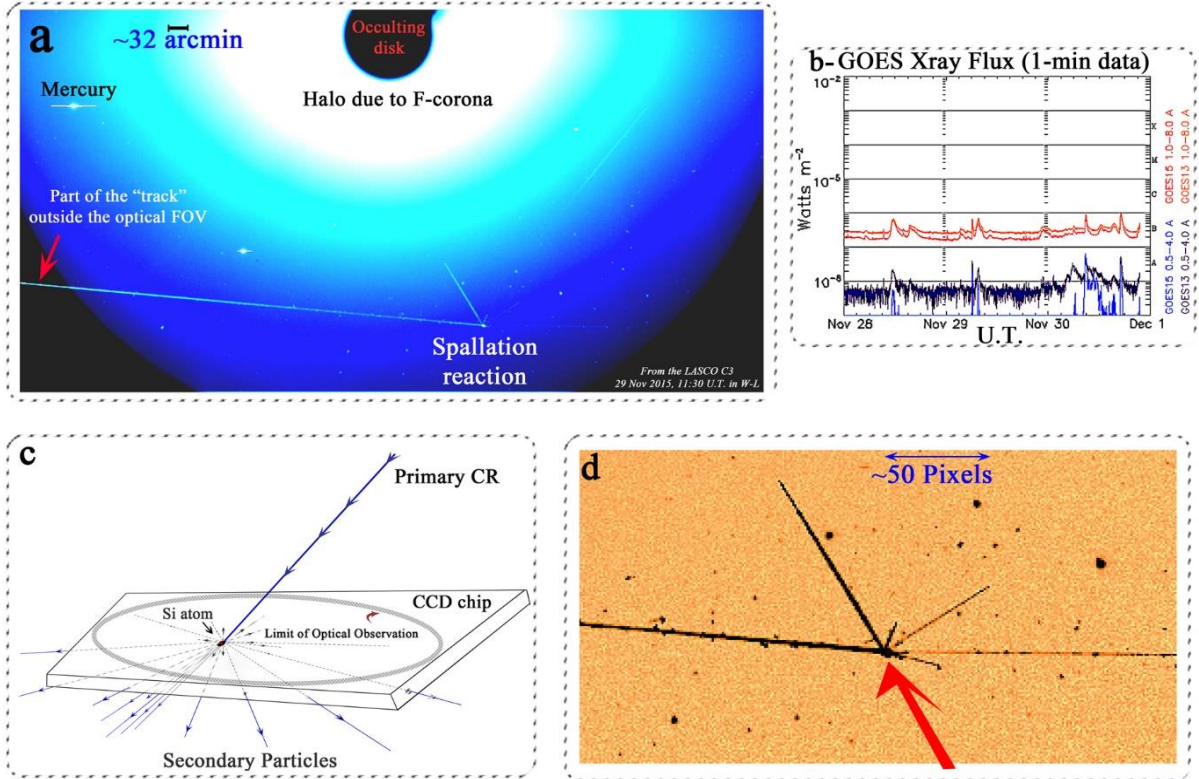

**Figure 3.** The original frame of the C3 coronagraph of LASCO *(SoHO* mission) taken at 11:30 UT on 29 Nov. 2015 at time of a quiet Sun with an overexposed image of the planet Mercury in the FOV. (a) Evidence of a spallation event shown at the bottom of the frame; reminding the first "stars" images from of galactic CR event recorded in balloon experiments (Levi Setti and Tomasini 1952). Note in panel a the dominating halo of the F-corona usually removed in routine images to enhance the K-corona variable structures; (b) the recording of the X-ray very low flux level radiation from the Sun given by the GOES satellite at the time of the observation, before and after; (c) schematic of the spallation event recorded inside the CCD chip following a collision by a high energy CR. The tracks of secondary particles produced at large angles will cross the chip and a small part will leave excited electrons that are subsequently read when the optical image is recorded. The tracks are not limited by the field of optical observations nor the external mask (d) the partial frame magnified in negative to show the region of the impact.

Here we show the original level 1 image to identify the whole event, including the part of the image outside the computer generated mask that usually is not shown on the processed images.

## 3 RESULTS

The important discovery-image is presented in Fig. 3. We note that the original image shows the tracks with a lower contrast than the routine processed image where several stationary components are accurately subtracted (after calibration), including the dominating F-corona (Koutchmy and Lamy 1985) in order to show the variable in time coronal structure (K-corona). In the bottom panel c) of Fig. 3 we show a contrast enhanced and magnified partial image. The linear features were enhanced using the Madmax operator (Koutchmy and Koutchmy 1989; Marshall et al. 2006; Tavabi et al. 2013) in the hope of finding the direction of the primary CR particle without great success. The panel b) of Fig. 3 presents a possibly realistic schematic with a suggested scenario, where tracks could be produced inside the thin chip.

This remarkable image, recorded in space, outside the Earth's magnetosphere, induced considerable discussion among the *SoHO* community concerned with the interpretation of C3 and C2 coronagraph images. After a quick-look analysis of LASCO images for several years (see Figs. 4 and 5) it appears that this event was probably the most energetic event observed for the more than 20 years of the *SoHO* mission, possibly a spallation event following a collision. Using the well-established power law of the spectral energy distribution of CRs observed on the ground (Freier et al. 1948; Dorman 2006; Kudela 2009), it is attempted to evaluate the range of maximum energy of CRs that will hit once the 400 mm$^2$ surface. We found the chip is of a 60 micron effective thickness in the LASCO camera and we follow the analysis performed in laboratories (*e.g.* Kowalczyk 2008; Putze 2009). Assuming a normal angle of arrival and a duration of 10 years and a maximum probability for the particle interacting at least once with a nucleus of Si of the chip during the crossing of the detector, our particle is found in the range of the $10^{13}$ to $10^{14}$ eV. Particles of this energy are considered to be produced within our Galaxy and are called GCRs.

This 1$^{st}$ evaluation is however not realistic because of the assumption of a signature every time the particle hits the chip.

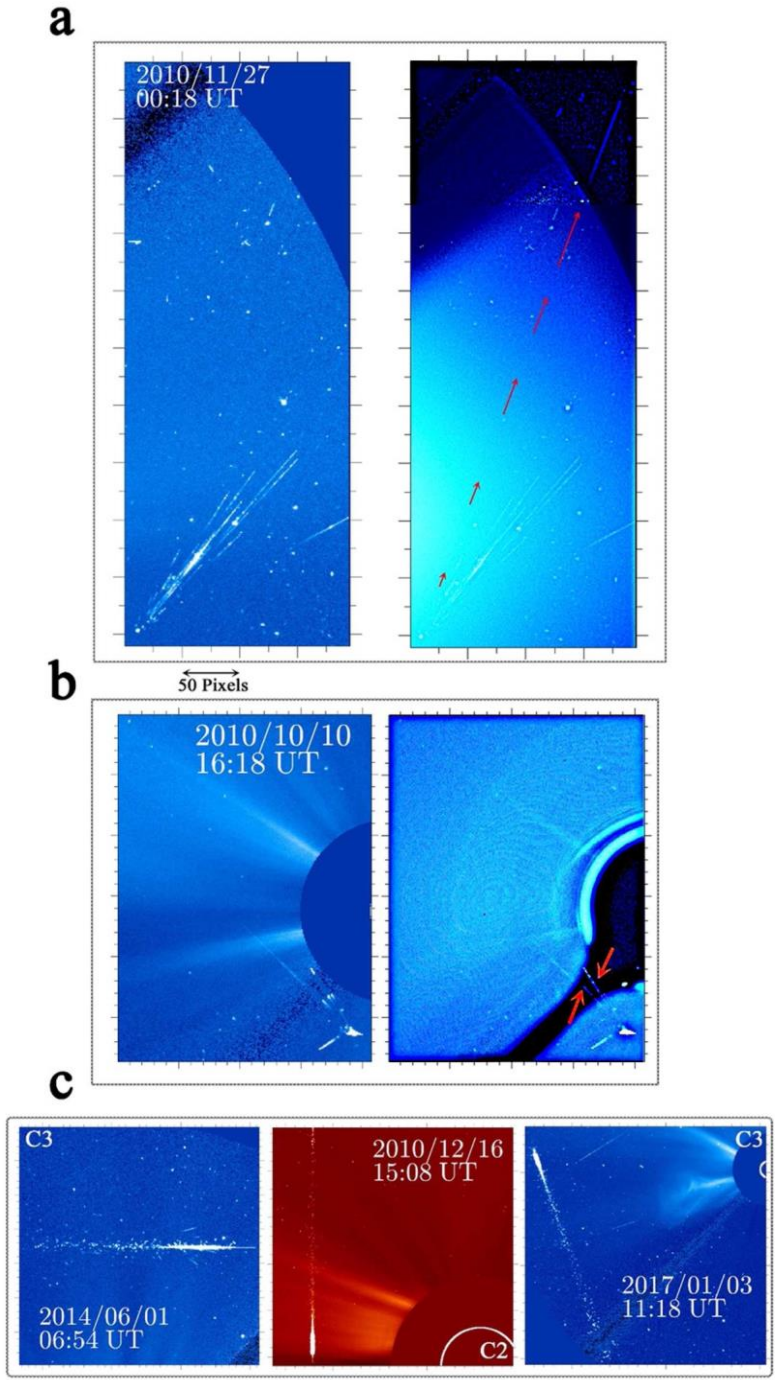

**Figure 4.** Example of different type of spallation events detected on single frames from the *SoHO* W-L coronagraphs database: a) linear tracks suggesting an impact with small pitch angle; note the part of a track seen only outside the optical FOV of the instrument (on the direction of the top red arrows); b) impact producing secondary particles inside the chip with large pitch angle; c) single thicker track suggesting heavier particles propagating inside the chip and producing some "ionization" effect around their track.

The equivalent nuclear cross section of the Si chip or equivalently, the mean free path of a particle crossing the silicon detector should be considered. We compared our detector sensitivity with what exists in the Earth's atmosphere (the nuclei of Oxygen and of Nitrogen are concerned) and use the measured proton-air inelastic cross-section measured by accelerators and cosmic-ray experiments for the range of energies between $10^9$ and $10^{13}$ eV from the Belov (2013) Fig. 2.

The value is typically 280 to 300 mb or $0.3\times10^{-28}$ m$^2$ for energies in the range of $10^9$ to $10^{13}$ eV. Further, the effective radius of the nucleus is $10^{-5}$ times the radius of the Si atom which is of order of 5 Å or $0.5\times10^{-9}$ m. For a proton it is $0.84\times10^{-15}$ m or 0.84 fm. The nucleus of Si is made of 14 protons and 14 neutrons. Note the cubic cell structure of the atom of Si with dimension 5.43 Å. The corresponding mean free path for a proton will be of order of 10 m.

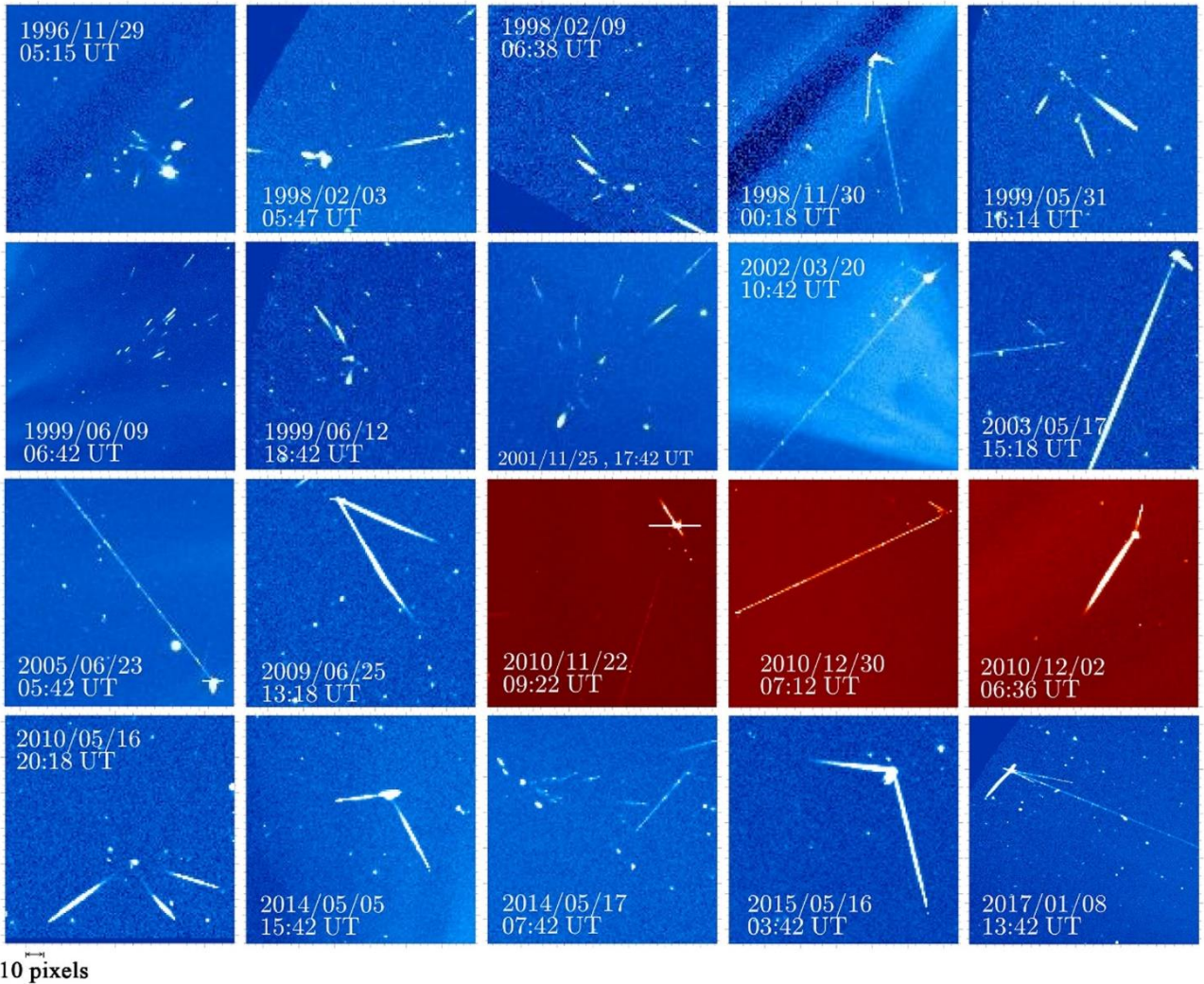

**Figure 5.** C2 & C3 selected spallation events recorded during the *SoHO* mission (1996 to 2017). Frames correspond to processed images after removing the stationary background as done at NRL. It is possible that this procedure favors the detection of the larger FOV observations taken with the C3 coronagraph (blue frames) compared to the C2 coronagraph images (red frames). Days with significant flares and/or CMEs on the Sun are carefully avoided.

Using these approximations we finally got a probable occurrence of a nuclear collision producing a spallation event in our chip of 20×20 $mm^2$ and 6×$10^{-5}$ m thickness for a period of 10 years when the flux is very close to 1 *part.*/sec/$m^{-2}$. The corresponding energy is then $10^{11}$ eV or 100 GeV which is definitely above what is measured at most for a solar CR (Miroshnichenko 2015), confirming our assumption of an event of GCR origin.

## 4 DISCUSSION AND CONCLUSIONS

Let us discuss several, presumably lower energy similar cases or cases not corresponding to a central collision that we found after using the whole set of LASCO *(SoHO* mission) coronagraph images from both C3 and C2 instruments. Fig. 4 presents 5 different types of events found in sequences recorded in 2010-2017. It gives a better perspective of effects (tracks) produced inside the chip by different CRs energies or different type of collisions. The most abundant CRs are made of protons but alpha particles ($He^{++}$) and heavier nuclei could also hit the chip (Schimmerling et al. 1996; Kudela 2009). In Fig. 5 we show selected events made from single processed frames found among the thousands sequences of full day coronagraphic observations, avoiding the case of SEPs at the same time as flares or CMEs. Note that all cases that we picked up show a single event with several related tracks, short and/or long. Again the case of Nov. 29, 2015 discussed above shows the best case showing both the location of the collision as a 2 px size "very bright point" with the secondary particles producing several divergent tracks of different length depending of the angle of their real tracks with respect to the plane of the chip, see Fig. 2. From a simple probabilistic evaluation, after evaluating the typical number of CR dots/frame taken at 12 minute cadence during periods outside solar activity events, we deduce an average maximum energy>1 TeV for particles responsible for the dots-events. However, an independent evaluation of the most probable value of energy of our best observed event of 29 Nov. 2015 (see Fig. 3) points to a range of energies of about $10^{11}$ eV. In addition, we do not have a comparable "star" signature for SEPs (Ramaty et al. 1996) occurring at the time of a big flare producing a large halo CME. We suggest that our events are of galactic origin of the most energetic particles that we record using the CCD imaging techniques and point out the consequences of this discovery made quite far from Earth.

The origin of the GCR particles is not known. We also tentatively looked at the temporal variations of the number of events observed at a one-month resolution. We note a strong variation in time depending of the level of the solar activity. This is well known from G-B observations and the so-called Forbush effect (Kudela 2009). The expulsion of solar magnetic clouds related to CMEs (Lara et al. 2005) and the quasi- stationary co-rotating interplanetary magnetic sectors deflect GCRs (see Fig. 1). This is also a part of the SEP variations that we avoid when counting the GCRs. It possibly produces some bias, partly explaining why we see a solar cycle variation much larger (a factor of 2 instead of 30%) than what we see on the ground.

Note that our data are taken well outside the Earth's magnetosphere significantly deflects SEPs and GCRs; it makes the records in GB data more affected. Accordingly, there is a suspicion that the modulation produced by solar activity on GCRs could be larger than what is usually given. Further we looked at the yearly modulation using the data for the full years 2000, 2008 and 2009. We found indications that the number of events per month is more numerous in December-January, at the years of minimum activity (2008 and 2009). Incidentally, this is the epoch when the *SoHO* spacecraft, which is always pointed towards the Sun with the CCD chip normal to that direction, also sees the center of our Galaxy. It is now believe that this region is made of a rotating massive black hole in Sagittarius and recently imaged (Ackermann et al. 2013; Cardillo et al. 2014) with Chandra. A lot of activity at the periphery is expected, including extreme magnetic activity accelerating elementary particles to GCR energy. Sources such as IC 443 and the Crab nebula SN remnant on the opposite direction of the sky (Grenier et al. 2015; Michael et al. 2016) could as well be significant. Finally, more GCRs could be determined using the whole set of available for more than 20 years observations and after some automatic method (Koutchmy and Koutchmy 1989; Marshall et al. 2006; Tavabi et al. 2013) of detecting the GCRs events in LASCO images is elaborated.

### ACKNOWLEDGEMENTS

The SoHO/LASCO data used here are produced by a consortium of the Naval Research Laboratory, Max-Planck-Institut fur Aeronomie (Germany), Laboratoire d'Astronomie de Marseille (France), and the University of Birmingham (UK). We are grateful to the SOHO/LASCO team for making their data publicly available. *SoHO* is a project of international cooperation between ESA and NASA. We warmly thank Hugh Hudson (Univ. of Glasgow) who was the first to encourage this research on CRs and who contributed very much in the discussion of the data and the presentation of this paper; Philippe Lamy (LAM- CNRS), Bernard Fleck (ESA), Russel Howard (NRL), Andrei Zhukov (ROB), Sergei Kuzin (LPI), Pierre Astier (IN2P3), Benoit Revenu (Nantes Univ), Leon Golub (SAO), Michel Dennefeld (IAP), Nicolas Prantzos (IAP), each brought their contribution to the discussion at different stages of the analysis. Francois Sevre (IAP) performed some additional analysis of the data and John Stefan (NJIT) diligently help with the manuscript; Guillaume Boileau performed a preliminary analysis of the data at IAP.

### REFERENCES

Ackermann, M., Ajello, M., Allafort, A., et al. 2013. Detection of the Characteristic Pion-Decay Signature in Supernova Remnants, Science, 339, 807-811

Baade, W. and Zwicky F., 1934. Remarks on Super-Novae and Cosmic Rays. Physical Review 46, 76-77

Belov, A. V., Bieber, J. W., Eroshenko, E. A., Evenson, P., Pyle, R., & Yanke, V. G. 2001, Proc. 27th Int. Cosmic-Ray Conf. (Hamburg), 9, 3507

Belov, K. 2013, Measuring chemical composition and particle cross- section of ultra-high energy cosmic rays by a ground radio array, in arXiv:1312.0382v1

Brueckner, G. E., et al. 1995, Sol. Phys., 162, 357

Butt, Y., 2009. "Beyond the Myth of the Supernova Remnant Origin of Cosmic Rays", Nature 460, 701-704

Cane, H. V., Richardson, I. G., & von Rosenvinge, T. T. 2010, JGR (Space Physics), 115, A08101

Cardillo, M.; Tavani, M.; Giuliani, A., 2014. The origin of Cosmic-Rays from SNRs: confirmations and challenges after the first direct proof, Nuclear Physics B (Proceedings Supplements), 256, 65-73

Chandrasekhar, S., and Fermi, E., 1953. Magnetic Fields in Spiral Arms. ApJ. 118, 113-115

Curdt, W. and Fleck, B., 2015. Solar and Galactic Cosmic rays observed by SOHO, Cent. Eur. Astrophys. Bull. 1, 1

Domingo, V., Fleck, B., & Poland, A.I., 1995. The SOHO Mission: an Overview. Sol. Phys., 162, 1-37

Dorman, L. ed., 2006. Cosmic Ray Interactions, Propagation, and Acceleration in Space Plasmas, vol. 339 of Astrophysics and Space Science Library

Firoz, K. A., Moon, Y. -J., Cho, K. -S., Hwang, J., Park, Y. D., Kudela, K., & Dorman, L. I., 2011. On the relationship between ground level enhancement and solar flare J. Geophys. Res., 116, A04101

Freier, P., Lofgren, E. J., & Oppenheimer, F., 1948. The Heavy Component of Primary Cosmic Rays, Phys. Rev., 79, 1818

Fermi, E., 1954. Galactic Magnetic Fields and the Origin of Cosmic Radiation, ApJ, 119, 1-6

Gaisser, T. K., 1990. nCosmic Rays and Particle Physics, Cambridge: Cambridge Univ. Press,

Grenier, I. A., Black, J. H., & Strong, A. W., 2015. The Nine Lives of Cosmic Rays in Galaxies, ARA&A, 53, 199-246

Kahler, S. W., 1992. Solar Flares and coronal mass ejections ARA&A 30, 113K

Klein, K.-L., Trottet, G., Lantos, P., & Delaboudini'ere, J.-P. 2001, A&A, 373, 1073

Koutchmy, S. and Lamy, P. L., 1985. Properties and Interactions of Inter planetary Dust, R. H. Giese & P. L. Lamy (Eds.), ASSL 119, 63

Koutchmy, O. and Koutchmy, S., 1989. Optimum filter and frame integration application to granulation pictures, in Proc. 10th Sacramento Peak Summer Workshop, High Spatial Resolution Solar Observations, ed. O. von derLuhe (Sunspot: NSO), 217

Kowalczyk, A. 2008, Proton induced spallation reaction in range 0.1-10 Gev, Ph.D Jagiellanian University, Cracow

Krasa, A. 2010, Spallation reaction Physics, coursec Czech Techn. Univ. Prague

Kudela, K., 2009. On energetic particles in space. Acta Phys. Slovaca, 59, 537-652

Marshall, S. Fletcher, L. and Hough, K., 2006. Optimal filtering of solar images using soft morphological processing techniques, Astron. Astrophys. 457, 729

Lara, A., Gopalswamy, N., Caballero-Lofpez, R. A., Yashiro, S., Xie, H., & Valdefs-Galicia, J. F. 2005, ApJ, 625, 441

Levi Setti, R. and Tomasini, G. 1952, Slow Heavy Mesons from Cosmic Ray Stars, Nuevo Cimento, Vol. IX, N. 12, 1242

Michael D. Delp, Jacqueline M. Charvat, Charles L. Limoli, Ruth K. Globus & Payal G., 2016.Apollo Lunar Astronauts Show Higher Cardiovascular Disease Mortality: Possible Deep Space Radiation Effects on the Vascular Endothelium, Nature Scientific Reports, 6:29901 DOI: 10.1038/srep29901

Miroshnichenko, L. 2015, Solar Cosmic Rays, Astrophys. And Sp. Sc. Library, 405, DOI 10.1007/978-3-319-09429-8

Mishev, A.; Usoskin, I. 2016, Solar Physics, 291, 1579-1580

Obayashi T., 1964. The Streaming of Solar Flare Particles and Plasma in Interplanetary Space, Space Science Reviews, 3, Issue 1, 79-108

Antje Putze. Phenomenologie et detection du rayonnement cosmique nucleate. Cosmologie et astrophysique extra-galactique [astro-ph.CO]. Universite Joseph-Fourier - Grenoble I, 2009. Francais.

Roederer, J. G., 1964.Generation, Propagation and Detection of Relativistic Solar Particles, Space Science 3, Issue 4,487-511

Ramaty, R., Mandzhavidze N., & Kozlovsky, B., 1996. in AIP Conf. Proc. 374, High Energy Solar Physics, ed. R. Ramaty, N. Mandzhavidze, & X.-M. Hua (New York: AIP), 172-180

Simpson, J.A., 1957. Cosmic Radiation Neutron Intensity Monitor. Annals of IGY IV, Pergamon Press, London p. 351

Schimmerling, W., J. W. Wilson, J. E. Nealy, S. A. Thibeault, F. A. Cucinotta, J. L. Schinn, M. Kim, and R. Kiefer, 1996. Shielding against Galactic cosmic rays, Adv. Sp. Res., 17, (2)31-(2)36

Smart, D. F. and Shea M. A., 2009. Fifty years of progress in geomagnetic cutoff rigidity determinations, Adv. Space Res. 44, 1107-1123,

Tavabi E., Koutchmy S., Ajabshirizadeh A., 2013. Increasing the Fine Structure Visibility of the Hinode SOT Ca II H Filtergrams, Sol. Phys. 283,187

Yagoda H., 1962. Radiation Studies in Space with Nuclear Emulsion Detectors, Space Science Reviews, 1, 224-277